\documentclass[]{spie}  

\usepackage{amsmath,amsfonts,amssymb}
\usepackage{subfig}
\usepackage{graphicx}
\usepackage[colorlinks=true, allcolors=blue]{hyperref}
\usepackage[ruled,vlined]{algorithm2e}
\graphicspath{{images/}}
\title{Statistically Adaptive Filtering for Low Signal Correction in X-ray Computed Tomography}

\author[a]{Obaidullah Rahman}
\author[a]{Ken D. Sauer}
\affil[a]{Dept. of Electrical Engineering, University of Notre Dame, IN, USA}
\author[b]{Charles A. Bouman}
\affil[b]{School of Electrical and Computer Engineering, Purdue University, IN, USA}
\author[c]{Roman Melnyk}
\author[c]{Brian Nett}
\affil[c]{General Electric Healthcare, Waukesha, WI, USA}

\authorinfo{Further author information: (Send correspondence to O.R.)
\\O.R.: E-mail: orahman@nd.edu, Telephone: 1 574 298 6896
\\K.D.S.: E-mail: sauer@nd.edu, Telephone: 1 574 631 6999
\\C.A.B.: E-mail: bouman@purdue.edu
\\R.M.: E-mail: roman.melnyk@ge.com
\\B.N.: E-mail: brian.nett@ge.com}

\begin{document} 
\maketitle

\begin{abstract}
Low x-ray dose is desirable in x-ray computed tomographic (CT) imaging due to health concerns. 
But low dose comes with a cost of low signal artifacts such as streaks and low frequency bias in the reconstruction. 
As a result, low signal correction is needed to help reduce artifacts while retaining relevant anatomical structures. 
 Low signal can be encountered in cases where sufficient number of photons do not reach the detector to have confidence in the recorded data.
X-ray photons, assumed to have Poisson distribution, have signal to noise ratio proportional to the dose, 
with poorer SNR in low signal areas. 
Electronic noise added by the data acquisition system further reduces the signal quality.

In this paper we will demonstrate a technique to combat low signal artifacts 
through adaptive filtration. 
It entails statistics-based filtering on the uncorrected data, correcting the lower signal areas more aggressively than the high signal ones. 
We look at local averages to decide how aggressive the filtering should be, and local standard deviation to decide how much detail preservation to apply. 
Implementation consists of a pre-correction step i.e. local linear minimum mean-squared error correction, followed by a variance stabilizing transform, and finally adaptive bilateral filtering. 
The coefficients of the bilateral filter are computed using local statistics.
Results show that improvements were made in terms of low frequency bias, streaks, local average and standard deviation, modulation transfer function and noise power spectrum.
\end{abstract}


\section{INTRODUCTION}
\label{sec:intro}  
Concerns over long-term health effects of x-ray
exposure in CT has moved the medical community in the direction of minimal dosage in clinical settings \cite{smith2009radiation} \cite{de2004risk}. 
As a result, the research motivation for CT dose reduction has grown lately under the industry-guiding principle of ALARA (as low as reasonably achievable).
The easiest way to lower the radiation dose is to reduce the x-ray flux by reducing the tube current and shortening the exposure time. 
However, simply lowering the radiation dose will, especially in presence of high attenuating material or large patients, severely degrade the image quality and diagnostic capabilities. 
To address this problem, low signal correction (LSC) and postprocessing algorithms have become necessary.

X-ray photon emission can be accurately modeled as a Poisson process, with SNR
proportional to the x-ray intensity. 
Therefore at low intensities of detected photons, photon counting noise may
overwhelm diagnostically important information. 
At the detector, the data acquisition system (DAS) adds electronic noise.
This has negligible effect in high signal data, but further damages SNR in low signal areas,
possibly driving registered counts below zero.

Sinogram domain correction \cite{kachelriess2001generalized} is generally preferred over image domain to
correct low signal artifacts because the low signal errors are more localized in the projection domain. 
Sinogram filtration includes techniques which could be as simple as a local averaging or 
Gaussian filtering, or could be treating with a custom-designed filter \cite{thibault2006recursive}. 
These filters adaptively correct the signal based on the signal or noise level.
CT vendors have their own specific filters that are designed to suit their customer preferences.

The goal of this research was to
provide a robust, adaptive filtering method to more fully exploit local statistical variation
in sinogram data than previous approaches.
Some of the image quality metrics that we will look at are degree of bias and streak correction, modulation
transfer function (MTF) profile, noise power spectrum (NPS), local sample average and standard deviation.

\section{Method}
\label{sec:Method}

We are attempting to reduce streaks, reduce low frequency bias, and reduce non-uniform texture 
while maintaining good resolution in the reconstruction image.
Counts data are approximately Poisson distributed and are independent,  conditioned on the integral projections. 
A well chosen filtering action can reduce the noise using some kind of local spatially weighted averaging.
Most LSC algorithms involve adaptive filters \cite{hsieh1998adaptive, kachelriess2001generalized}, or non-linear sinogram filtering. 
Bilateral filters have also shown promise in correcting low signal artifacts \cite{manduca2009projection}.
We seek a solution that entails signal-adaptive filtering action in applying bilateral filters
in the domain of a variance stabilizing transform.

The CT imaging chain starts with scanning of patients and acquisition of photon counts.
It is followed by correction of low signal artifacts, conversion of 
counts to integral projection estimates by the $-\log()$ operator
and, typically, FBP reconstruction and possibly image-domain post-processing.
Our algorithm, adaptive filering low signal correction (AF LSC), deals with the low signal correction
part of the chain and comprises the following components.

\subsubsection{Local linear minimum mean-squared error (LLMMSE) correction}
Before the negative log step converting photon counts to integral projections, 
the counts need to be positive. 
In cases of near-complete photon starvation, large numbers of negative registered
counts pose a unique problem.
Simply forcing the negative data to non-negative values creates a bias in the data, 
so we wish to correct them with minimal disturbance of the local mean.
The electronic noise in this case injects a large fraction of the variance.
Since this noise is independent of the signal, its variance can be estimated before the scan.
Its statistics can be captured by recording detector response with the x-ray beam turned off.
It is assumed to be independent Gaussian\cite{xu2009electronic} \cite{ma2012variance}. 
The local, linear, minimum mean-squared error (LLMMSE) filter \cite{chang2016modeling}
removes a large fraction of negative data points through adaptive linear filtering.
LLMMSE is a pre-correction step, and we limit it to only very low values of SNR.

\subsubsection{Variance stabilizing transform (VST)}
Photon count levels in common clinical scans vary by several orders of magnitude,
with large offsets possible in sinogram values in the distance of a few detectors
in the presence of heavily attenuating materials.
Because there exists a large number of well-designed algorithms in the literature to tackle 
constant-varaince Gaussian noise, a key step in our algorithm is transformation to 
approximately constant variance in the data exiting the LLMMSE step.
The VST transforms a Poisson random variable to have a variance that does not depend on its mean \cite{anscombe1948transformation}.
Once the data has approximately constant variance, it is similar to Gaussian and therefore
traditional Gaussian denoisers can be applied to it.
In the paper \cite{makitalo2012optimal} authors show how to perform the VST when the 
data is corrupted with Poisson-Gaussian noise. 
This work also provides a closed form unbiased inverse VST.

\subsubsection{Bilateral filtering}
Some existing LSC methods choose among a fixed bank of filters based on the measured counts. 
This piece-wise approach is somewhat restrictive, so we proposed an adaptive approach in which the filter parameters depend on the local signal and noise level.
Bilateral filtering in the VST-transformed counts is the heart of our LSC algorithm.
The bilateral transform has been shown to work well with images \cite{zhang2008multiresolution},
and the authors in \cite{manduca2009projection} have implemented it in projection space.
The parameters the authors used in this case were fixed and not dependent on local statistics.
Since the recorded counts can have a large dynamic range, choosing one set of parameters for the entire data set is a limitation in responding to serious low signal problems. 
We would like to do extremely aggressive correction in the counts at comparable levels to 
the standard deviation of the electronic noise, and less for higher counts. 
This is the principle innovation in our adaptive filtering.
The authors in \cite{lee2019deep} have also implemented adaptive bilateral filtering but only the spatial term, and they
compute the coefficients of the filter using a neural net.

The bilateral filter outputs a weighted sum of a datum and its neighbors for each data point. 
Each weight is decided by
the respective neighbor's spatial proximity to the current datapoint and the value of the neighboring measurement.
The first part (i.e. spatial term) in the weight decreases as the spatial distance of the 
neighbor pixel w.r.t the center or current voxel, as in a conventional filter.
The second part (intensity term) relaxes the degree of filtering in the case of
large difference in intensity between the neighbor and the current pixel.
If the distance and difference are high, the neighbor has decreased weight
and filtering tends to be done separately on the two sides of an edge.

\subsubsection{Inverse VST and positivity mapping}
After adaptive, biliateral filtering, we
convert the data from the VST domain back to counts domain for subsequent processing. 
An unbiased VST inverse is implemented as explained in \cite{makitalo2012optimal}. 
After the inverse VST, there could be a small number of zero values which need to be mapped to positive numbers before the negative log step.
We effect this with a simple exponential mapping for any value below a set threshold.
\\ \\ The algorithm in summarized in Algorithm \ref{algo: BF algorithm}.

\begin{algorithm}[]
\SetAlgoLined
 Get raw counts from the reconstruction chain\;
  $\lambda \gets \text{raw counts}$\ $\sigma_e \gets \text{standard dev. of electronic noise}$\ ;
  $N \gets \text{number of data points in raw counts array}$\;
  $\lambda_{av} \gets \text{local average of raw counts}$\;
  $\eta = \frac{\lambda_{av}}{\lambda_{av} + \sigma_e^2}$\;
  $\lambda_{th} \gets \text{A scalar, counts below which undergo LLMMSE correction}$\;
  $\lambda_{th}' \gets \text{A scalar, counts below which undergo positive mapping}$\;
  \If{$\lambda \leq \lambda_{th}$}
  	{
   		$\lambda_{llmmse} = \eta \lambda + (1-\eta)\lambda_{av}$\;
   	}
  $\lambda_{vst} = 2\sqrt{\lambda_{llmmse} + \frac{3}{8}}$\;
  \For{$i \gets 1 \text{ to } N$}
  	{
   		\For{$j \in \Omega_i$}
   		{
   		$W_j = e^{-\frac{|i-j|}{\sigma_d}} e^{-\frac{|\lambda_{vst}^{(i)}-\lambda_{vst}^{(j)}|}{\sigma_r}}$\;
   		}
   		$\lambda_{bf}^{(i)} = \frac{\sum_{j \in \Omega_i} W_j \lambda_{vst}^{(j)}}{ \sum_{j \in \Omega_i} W_j }$\;
   	}
   	$\lambda_{ivst} = \frac{1}{4}(\lambda_{bf})^2 + \frac{1}{4}\sqrt{\frac{3}{2}}(\lambda_{bf})^{-1} - \frac{11}{8}(\lambda_{bf})^{-2} + \frac{5}{8}\sqrt{\frac{3}{2}}(\lambda_{bf})^{-3} - \frac{1}{8}$\;
   	$\hat{\lambda} = \lambda_{th}' e^{\frac{\lambda_{ivst}}{\lambda_{th}'}-1} \quad \quad \forall \ \lambda_{ivst} < \lambda_{th}'$\;
   	Plug $\hat{\lambda}$ back into the reconstruction chain\;
 \caption{Adaptive filtering (AF) LSC algorithm \label{algo: BF algorithm}}
\end{algorithm}


\newpage
\section{Results}
\label{sec:Results}
To get $\sigma_d$, and $\sigma_r$, local average and standard deviation, in a 
sinogram window of $7\times 5\times 3$ were computed for the array of received photon counts, $\lambda$, 
which resulted in arrays $\hat{\mu}$ and $\hat{\sigma}$ respectively.
For $i^{th}$ datapoint, $\sigma_d^{(i)} = K_1\frac{1}{\hat{\mu}^{(i)}}$, and  $\sigma_r^{(i)} = K_2\hat{\sigma}^{(i)}$, where $K_1 = 400,\ K_2 = 5$. Bilateral filtering was performed in a 3-D window of $Channel \times Row \times View = 13 \times 7 \times 3$.
\begin{figure}[]
  \begin{center}
	\subfloat[]{\includegraphics[scale=0.35]{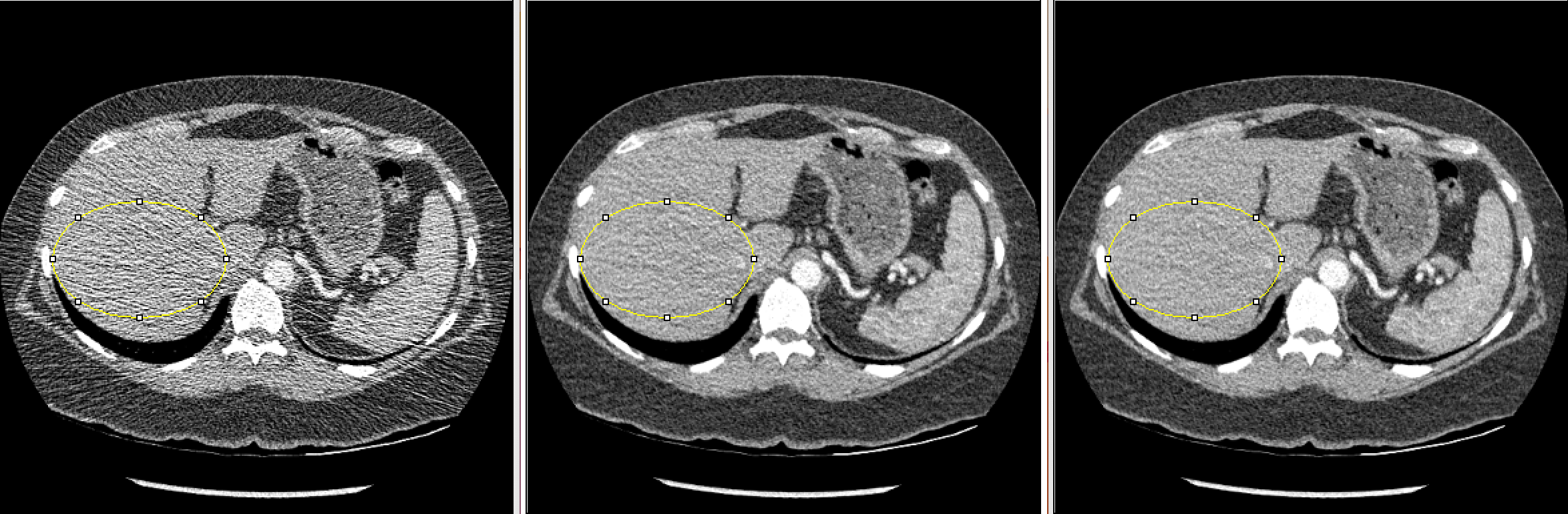}\label{fig:Exam1_70_imagej}}
	\hspace{0.1 cm}
	\subfloat[]{\includegraphics[scale=0.35]{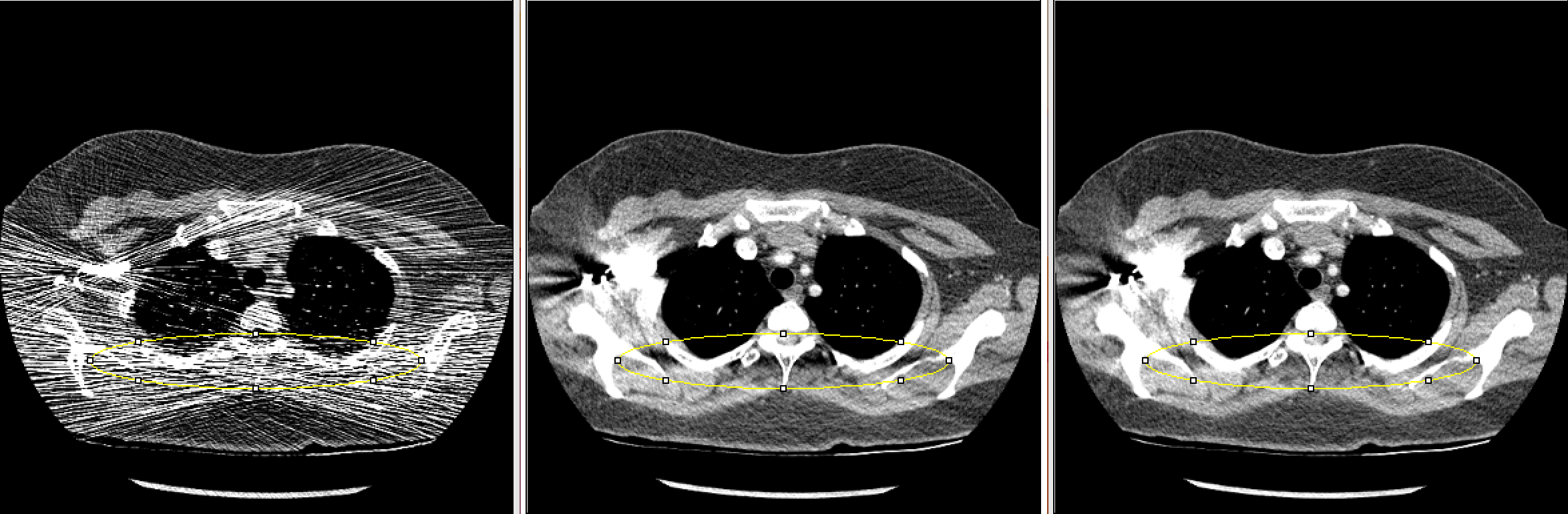}\label{fig:Exam1_390_imagej}}
	\hspace{0.1 cm}
	\subfloat[]{\includegraphics[scale=0.375]{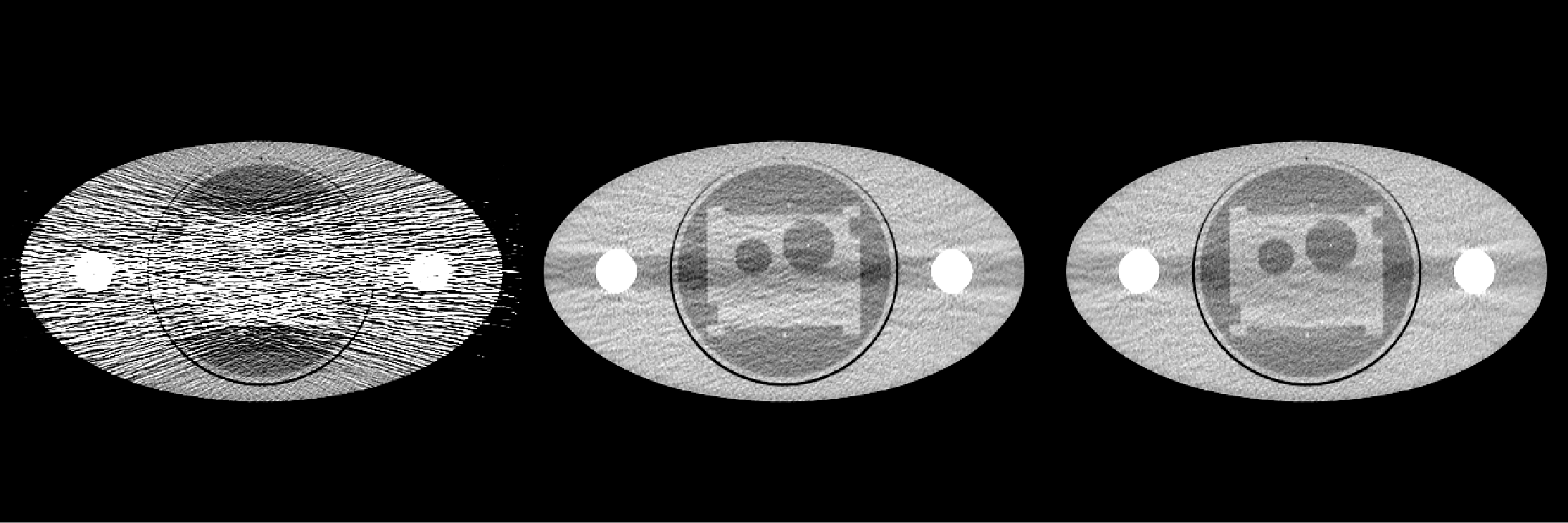}\label{fig:Exam6_10}}
	\caption{A slice of reconstructed image (left) Uncorrected, (center) FT LSC, (right) AF LSC
	\label{fig:correction recon images}
	\protect\subref{fig:Exam1_70_imagej} An axial slice of clinical image reconstruction in the liver region \protect\subref{fig:Exam1_390_imagej} An axial slice of clinical image reconstruction in the shoulder region containing contrast \protect\subref{fig:Exam6_10} An axial slice of low signal phantom image reconstruction}
  \end{center}
\end{figure}

For a practical comparison, we use an LSC algorithm
similar to\ \cite{hayes2018low} which employs filtering based on lower and upper thresholds.
The data points valued less than the lower threshold undergo full box-car filtering,
and the data points higher than the upper threshold undergo median filtering.
This or its variants are typical examples of simple yet effective low signal correction techniques used commercially.
We will label it simply ``fixed threshold" (FT) LSC for results below.

It is clear that streaks are reduced in the FT LSC as well as AF LSC images.
Compared with FT LSC, in AF LSC local averages in the reconstruction images were preserved, with slight reduction in local variance.
In Fig. \ref{fig:Exam1_70_imagej}, the liver region in AF LSC looks more uniform.
In Fig. \ref{fig:Exam1_390_imagej}, we see a reduction in low frequency bias around and below the spine.

\begin{table}
	\begin{minipage}{0.5\linewidth}
	\centering
		\caption{MTF measurement on FT LSC and adaptive filtering LSC}
		\label{tab:MTF}
		\begin{tabular}{|c|c|c|}
			\hline
			{\bf MTF (w.r.t. DC)} & \multicolumn{2}{c|}{{\bf Frequency ($cm^{-1}$)}}\\
			\hline
			& {\bf FT LSC} & {\bf AF LSC}\\
			\hline
			$50\%$ & 1.61 & 2.14\\
			\hline
			$10\%$ & 5.32 & 6.29\\
			\hline
			$4\%$ & 7.37 & 8.27\\
			\hline
		\end{tabular}
	\end{minipage}\hfill
	\begin{minipage}{0.5\linewidth}
		\centering
		{\includegraphics[scale=0.50]{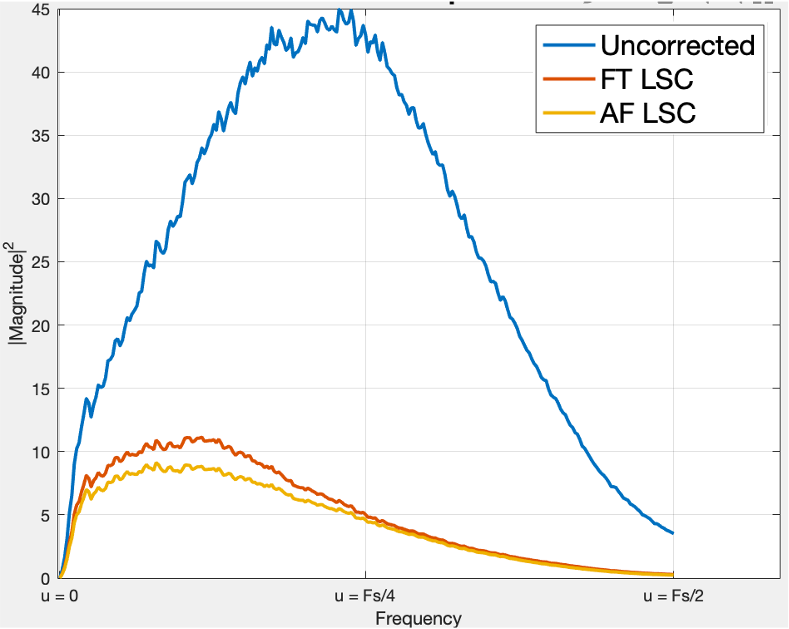}}
		\captionof{figure}{Radially averaged profile plot of noise power spectrum in the liver region \label{fig:Radial_NPS}}
	\end{minipage}
\end{table}

MTF measurement was made in a region of interest containing a titanium wire in a low signal phantom. 
Measurements are rendered for the frequencies at which the response falls to 50\%,
10\% and 4\% of maximum.
It can be seen in Table \ref{tab:MTF} AF LSC receives improved MTF scores at each of these points. 
The noise power spectrum for AF LSC is flatter than that of FT LSC and uncorrected images, as seen in Fig. \ref{fig:Radial_NPS}.
This property is typically viewed as advantageous for image evaluation.
From the NPS plot it can also be inferred that AF LSC has lower variance in liver region compared to FT LSC.

\section{Conclusions}
Adaptive filter LSC is a relatively simple technique, but is highly adaptive to signal and noise levels
in CT sinograms with very low photon counts.
In early evaluations, it appears to offer several improvements over
FT methods: reduced streaks and low frequency bias, 
improved spectral properties of noise,  
more uniform texture, 
better resolution as measured by MTF score,
and lower standard deviation within uniform ROIs. 

\bibliography{bibliography} 
\bibliographystyle{spiebib} 

\end{document}